\documentstyle[12pt,epsf,oldlfont,amsfonts]{article}

\expandafter \ifx \csname NoFigures\endcsname\relax\else
	\def\epsfbox#1{PostScript(TM) figure {\tt#1} to appear here.}
\fi

\title{
	The abstract boundary---a new approach to singularities of manifolds.%
	\protect\footnotetext{
		To appear in {\em J.~Geometry~Phys.}  13:223--253 (1994).
	}
}
\author{Susan M. Scott\thanks{Centre for Mathematics and its Applications, The
Australian National University, Canberra ACT 0200, AUSTRALIA, e-mail:
Susan.Scott@anu.edu.au} \thanks{ARC Australian Research Fellow}  \and Peter
Szekeres\thanks{Department of Physics and Mathematical Physics, The University
of Adelaide, Adelaide SA 5005, AUSTRALIA, e-mail:
pszekere@physics.adelaide.edu.au}}
\date{26 May 1994\\
	CMA Maths. Research Report No.~MRR028-94\\
	{\tt gr-qc/9405063}}

\newtheorem{definition}{Definition}
\newtheorem{theorem}[definition]{Theorem}
\newtheorem{example}[definition]{Example}
\newtheorem{examples}[definition]{Examples}

\newcommand{\qed}{\par\vspace{0.02em}\hfill
	\rule{0.65em}{0.65em}\vspace{0.04em}}
\newcommand{\lra}{\longrightarrow}
\newcommand{\ra}{\rightarrow}
\newcommand{\calc}{\mbox{${\cal C}$}}
\newcommand{\calm}{\mbox{${\cal M}$}}
\newcommand{\calu}{\mbox{${\cal U}$}}

\newcommand{\hM}{\mbox{$\widehat{\cal M}$}}
\newcommand{\bM}{\mbox{${\cal M}$}\kern-0.82em\rule[1.9ex]{0.72em}{.07mm}}
\newcommand{\vsp}{\vspace{2mm}}
\newcommand{\cbp}{${\cal C}$-boundary point}
\newcommand{\cbps}
  {${\cal C}$-boundary points}
\newcommand{\bm}{\mbox{${\cal B}(\cal M)$}}
\newcommand{\mb}{\makebox(0,0)}
\newcommand{\Rdot}[1]{\dot{\Bbb R}\raise1ex\hbox{$\scriptstyle #1$}}

\begin{document}

\maketitle
\begin{abstract}
A new scheme is proposed for dealing with the problem of singularities in
General Relativity. The proposal is, however, much more general than this. It
can be used to deal with
manifolds of any dimension which are endowed with nothing more than
an affine connection, and requires a family
\calc\ of curves satisfying a {\em bounded parameter property} to be
specified at the outset. All affinely parametrised geodesics are
usually included in this family, but different choices of
family \calc\ will in general lead to different singularity structures.
Our key notion is the {\em abstract
boundary\/} or {\em $a$-boundary\/} of a manifold, which is defined for any
manifold \calm\ and is independent of both the affine
connection and the chosen family \calc\ of curves. The $a$-boundary is
made up of equivalence classes of boundary points of \calm\ in all possible
open
embeddings. It is shown that for a pseudo-Riemannian manifold $(\calm,g)$ with
a specified family \calc\ of curves, the abstract boundary points can then be
split up into four main categories---regular, points at infinity,
unapproachable
points and singularities. Precise definitions are also provided for the notions
of a {\em removable singularity} and a {\em directional singularity}. The
pseudo-Riemannian manifold will be said to be singularity-free if its abstract
boundary contains no singularities. The scheme passes a number of tests
required of any theory of singularities. For instance, it is shown that all
compact manifolds are singularity-free, irrespective of the metric and chosen
family \calc. All geodesically complete pseudo-Riemannian manifolds are also
singularity-free if the family \calc\ simply consists of all affinely
parametrised geodesics. Furthermore, if any closed region is excised from a
singularity-free manifold then the resulting manifold is still
singularity-free.
Numerous examples are given throughout the text. Problematic cases posed by
Geroch and Misner are discussed in the context of the $a$-boundary and are
shown
to be readily accommodated. \end{abstract}

\section{Introduction}
In general relativity one often wishes to know whether a particular solution of
 Einstein's field equations is singular or not. Such a seemingly simple
question has frequently been the cause of a great deal of confusion. The most
common problem is that a solution usually comes packaged in one of two ways.
Either it is embedded in a larger four-dimensional manifold (e.g.\ the
Schwarzschild solution $r>2m$) or no embedding is given at all (e.g.\
Minkowski space in its usual coordinates). In the latter case there is no edge
to the space--time, making it difficult to assess where any singular behaviour
might occur. In the former case the metric may look singular with respect
to the particular embedding given, but may not look singular at all with
respect
to another embedding (e.g. Kruskal's embedding for the Schwarzschild
solution \cite{Krusk}).

Historically there have been several approaches to this problem. Starting with
the work of G. Szekeres \cite{GSzek}, who was probably the first to discuss the
importance of geodesic completeness, the next decade saw several attempts
to provide explicit boundary constructions for
space--times. The most important of these were Geroch's $g$-boundary
\cite{Ger}, the causal or $c$-boundary of Geroch, Kronheimer and Penrose
\cite{GKP} and Schmidt's  $b$-boundary \cite{Schmidt}. Excellent reviews of
the situation, up till about 1977, can be found in refs.
\cite{HE,ShepRy,ES,CS}.
Since that time there has been little advancement in this field.

Each of the constructions mentioned above, however, suffers from
various problems and limitations, perhaps the worst being the difficulty of
applying them to specific examples.
For instance, the Schmidt construction involves studying the 20-dimensional
bundle of frames for the given manifold, a daunting task to say the least. In
the few cases where it has been possible to compute the boundary explicitly
(in particular the two-dimensional Friedmann model) the results have not been
encouraging from the physical point of view \cite{Bos}.

Our approach is motivated by a number of considerations. Firstly, we
want a definition of ``singularity'' which can be used in a reasonably
straightforward way on the sorts of examples that commonly arise in
general relativity. Secondly, in these examples, one often has an
intuitive or physical feeling for the structure of the singularity in question.
For example, the singularity at $r=0$ in the Schwarzschild solution
seems to be a spacelike hypersurface, while the studies of the Curzon
singularity interpret it as being generated by the world lines of
points on an infinitely large ring
\cite{SzekMorg,SzekScott1,SzekScott2}.
Our aim is to see if these intuitive notions can be
grounded in a more rigorous mathematical procedure.

Thirdly, we feel that too much has been made of the
differences between the positive definite case (Riemannian) and the
space--time case (Lorentz\-ian). The former, it is true, does have a well
posed theory of singularities via the Cauchy completion \cite{HE,Helgason},
but once the door is opened to more general metrics
it is hard to see why one would want to restrict attention to the case
of Lorentz\-ian signature. Indeed, as attention is focussed almost exclusively
on the behaviour of the geodesics, there should be a theory of singularities
which needs nothing more than an affine connection. A satisfactory singularity
theory of this kind could then accommodate other interesting theories such as
Einstein--Cartan, Kaluza--Klein, Yang--Mills, etc.

Within general relativity many discussions concentrate only on timelike or
causal geodesics, as though spacelike geodesics were of no physical
consequence.
This seems to be a very shortsighted point of view. In two dimensions, for
example, it is purely a matter of interpretation as to which dimension is
taken as ``space'' and which as ``time''. Also
if the curvature becomes infinite at some point which cannot be
approached by causal geodesics
but which can be approached by spacelike geodesics
(this nearly happens in the Reissner--Nordstrom solution),
then surely
the space--time should not be called ``singularity-free'', since there is an
obstruction to continuing certain geometrically important curves.

The key to singularity theory is the concept of an extension of a manifold.
Suppose one is given a ``boundary point'' $p$ of a manifold, arising
for example on the boundary of a coordinate patch used in the presentation of
a space--time
$(\calm,g)$. If \calm\ is continued through $p$ by making it part of a
larger manifold
\hM\ in which $p$ is covered by a new coordinate patch and the metric extends
to a metric $\hat{g}$ on \hM, then $p$ is clearly a regular boundary point. If
no such extension exists, however, a further possibility presents itself---it
may be a ``point at infinity'', unattainable by any geodesics with finite
affine parameter. If neither of these conditions apply, i.e., $p$ is
approachable by geodesics with finite affine parameter yet no extension of
\calm\ exists through $p$, then we have what we would term a ``singularity''.

This all needs to be made more precise, but basically singularities can be
thought of simply as ``failed'' boundary points of open embeddings of a
space--time $(\calm,g)$---i.e.\ points which are neither regular nor points at
infinity. For some people this may seem too narrow a concept, since our
boundary points always belong to open embeddings. We believe this
to be an adequate constraint however. Certainly regular boundary points are
always of
this type and so it is natural to classify as ``singular'' all such boundary
points which are not regular.
In this sense points at infinity are also singular,
but we will discard them as being ``infinitely far away''.

Our procedure will be to provide a series of precise definitions leading up
to the concept of singularity.
Many of the terms defined will be appearing for the first time or may have
appeared earlier in a different context. We have tried to choose words
which are as suggestive as possible of their meaning and liberally sprinkle
the text with examples which should clarify the need for the
various stratagems adopted in our definitions. Most theorems have short
proofs and are needed only to proceed to the next stage of the definitional
ladder, leading eventually to the concept of a singularity.

While space--times are clearly our main objective, more general
pseudo-Riemannian manifolds, or even manifolds with just an affine connection
will fall under our scheme. Thus from the outset we try to define classes of
curves which are ``geodesic-like'' on a manifold. This is done in section~2.
Classes of curves whose parameters have a property similar to that possessed by
affinely parametrised geodesics will be said to have the {\em bounded
parameter property\/} (b.p.p.\ for short). This notion is what will ultimately
be needed to test whether a boundary point is ``at infinity'' or not.
Without such a class of curves singularity theory has no meaning, since
boundary points can always be ``sent to infinity'' by an appropriate
change of parameter. Our scheme however is very general and
permits discussion of singularities with respect to many different classes of
curves such as causal geodesics, smooth curves with generalised affine
parameter, etc.

Central to our discussion is the notion of an open embedding, i.e.\ an
embedding $\phi$ of a manifold \calm\ in another manifold \hM\ of the
same dimension. It is used so often in this paper that we prefer to give
it the special name {\em envelopment}. Section~3 introduces the idea of
{\em boundary points} of an envelopment and develops
the key concept of one boundary point
{\em covering\/} another one (these belonging, in general, to different
envelopments).
  Basically $p$ covers $q$ if whenever one approaches $q$
(from within \calm) then one approaches $p$.  In a sense this says that $q$
can be thought of as being a
``part of'' $p$. Boundary points are then said to be equivalent
if they mutually cover each other, the equivalence classes defined by
this relation being called {\em abstract boundary points}.

In section~4 the special case of pseudo-Riemannian manifolds (any signature
metric) is discussed. There is nothing in this section which could not be
generalised to manifolds having just an affine connection.
The concept of an {\em extension\/} of a pseudo-Riemannian manifold
and, successively, the concepts
of {\em regular boundary points}, {\em points at infinity\/} and
{\em singularities\/} are discussed. We define rigorously what it means for a
point at infinity or a singularity to be {\em removable\/} or {\em essential}.
The latter concept is shown to pass to the abstract boundary.

In section~5 a complete
classification of boundary points and the abstract boundary is given, including
all possible ways in which different types of boundary points can cover each
other.

Section~6 is devoted to the problem of singularities. In particular, it is
clearly stated what it means for a pseudo-Riemannian manifold to be
{\em singularity-free}. Several theorems are derived, giving criteria for
a manifold to be singularity-free. Geroch's and Misner's problematic
examples are both discussed and shown to have unequivocal interpretations
in our scheme.

In section~7 we summarise the situation and focus on a number of unanswered
problems arising out of this paper.

A preliminary version of these ideas has been presented by one of us (S.M.S.)
\cite{Scott}.  Some concepts were not optimally developed at that stage and
the present version should be regarded as superseding the one given
there.  It is, however, a useful source of further examples to illustrate our
techniques.

\section{Parametrised curves}
In the following definitions we will always assume that \calm, \hM,
\bM, etc.\  refer to paracompact, connected, Hausdorff,
$C^{\infty}$-manifolds all
having the same dimension $n$. Unless specifically stated otherwise, it is {\em
not\/} assumed in this section or in section 3 which follows that the manifold
is endowed with a metric or affine connection.

\begin{definition} {\em By a {\em (parametrised) curve\/} in a manifold
\calm\ we shall mean a $C^{1}$ map $\gamma: I \ra \calm$ where $I$ is a
half-open
interval $[a,b), \;a<b\leq
 \infty$, whose tangent vector $\dot{\gamma}$ nowhere vanishes on this
interval.
Such a curve will be said to {\em start\/} from  $p=\gamma (a)$, and
the parameter will be said to be {\em bounded\/} if $b<\infty$,
{\em unbounded\/} if $b=\infty$.} \end{definition}

\begin{definition} {\em A curve $\gamma':[a',b') \ra \calm\ $is a {\em
subcurve\/}
of $\gamma$ if $a \leq a' < b' \leq b$ and $\gamma' = \gamma |_{[a',b')}$,
i.e., $\gamma'$ is the restriction of $\gamma$ to a subinterval.
If $a=a'$ and $b>b'$ then we say $\gamma$ is an {\em extension\/} of
$\gamma'$.}
\end{definition}

\begin{definition} {\em A {\em change of parameter\/} is a monotone increasing
$C^{1}$ function \[ s: [a,b) \lra [a',b') = I' \]
such that $s(a) = a', s(b)=b'$ and d$s/$d$t > 0$ for $t \in [a,b)$.
We say the parametrised curve $\gamma':I'\ra \calm$ is
{\em obtained from $\gamma$ by the change of parameter $s$\/} if
\[\gamma' \circ s = \gamma. \]} \end{definition}

\begin{definition} {\em Let $\cal C$ be a family of parametrised curves in
\calm\
such that \begin{quote}
(i) for any point $p \in \calm\ $there is at least one curve $\gamma$ of the
family  passing through $p$, \\
(ii) if $\gamma$ is a curve of the family then so is every subcurve of
$\gamma$, and\\
(iii) for any pair
of curves $\gamma$ and $\gamma'$ in \calc\ which are obtained from each other
by a change of parameter we have either that the parameter on both curves is
bounded or it is unbounded on both curves.
\end{quote}
Any family \calc\ satisfying conditions (i), (ii) and (iii) will be said to
have
the {\em bounded parameter property\/} (b.p.p.).} \end{definition}
\begin{examples} The following families of curves all have the b.p.p.
\\[2mm] (i)  Geodesics with affine parameter in a manifold \calm\ with affine
connection. A change of parameter must have the form $s=At+B$ and the bounded
parameter property is clearly satisfied. We denote this family by
$\calc_{g}(\calm)$. The term ``geodesic'' here always refers to a geodesic arc
starting from some point $p\in \calm$.
\\[2mm](ii) $C^1$ curves with generalised affine parameter \cite{HE}
in a manifold \calm\ having affine connection. This family will be
denoted $\calc_{gap}(\calm)$.
\\[2mm](iii) Timelike geodesics with proper time parameter in a Lorentzian
manifold \calm, denoted $\calc_{gt}(\calm)$. If the manifold is time-orientable
one can also talk of future-directed and past-directed timelike geodesics,
$\calc_{gt+}(\calm)$ and $\calc_{gt-}(\calm)$.
\label{bpp exs}
\end{examples}

\begin{definition} {\em We say $p\in \calm$ is a {\em limit point\/} of a curve
$\gamma:[a,b)\ra \calm$ if there exists an increasing infinite sequence of
real numbers $t_{i} \ra b$ such that $\gamma (t_{i}) \ra p$.}
\end{definition} An equivalent statement of this definition is to say that
for every subcurve $\gamma' = \gamma |_{[a',b)}$ of $\gamma$, where $a \leq
a' < b$,
$\gamma'(t)$ enters every open neighbourhood \calu\ of $p$.
\begin{definition} {\em
We say $p$ is an {\em endpoint\/} of the curve $\gamma$ if  $\gamma(t) \ra p$
as
$t\ra b$.} \end{definition} For Hausdorff manifolds this implies that $p$ is
the
unique limit point of $\gamma$.

\begin{definition} {\em Given a manifold \calm\ with a family \calc\ of
curves having the b.p.p.,
we say the manifold \calm\ is {\em \calc-complete\/} if every curve $\gamma
\in \calc$  with
bounded parameter has an endpoint in \calm.} \end{definition}

Of course this does not guarantee that every curve of
the family \calc\ with bounded parameter has an extension to a curve in
\calc. However, the converse {\em is} true, since by continuity, every curve
$\gamma:[a,b)\ra \calm$ which has an extension $\gamma':[a,b')\ra \calm$
($b'>b$), where $\gamma' |_{[a,b)} = \gamma$, clearly has $p=\gamma'(b)$ as
its endpoint.

In most practical cases such as families of geodesics with affine parameter,
extendability of all curves with bounded parameter and \calc-completeness are
equivalent.

\section{Enveloped manifolds and boundaries}
\begin{definition} {\em An {\em enveloped manifold\/} is a triple
$(\calm,\hM,\phi)$
where \calm\ and $\hM$ are differentiable manifolds of the same dimension $n$
 and $\phi$ is a $C^{\infty}$ embedding
$\phi:\calm\ra \hM$.} \end{definition}

Since both manifolds have the same dimension $n$, $\phi(\calm)$ is an open
submanifold of
\hM. \calm\ is often identified with  $\phi(\calm)$ in the natural way,
when there is no
risk of ambiguity. The enveloped manifold will also be referred to as an {\em
envelopment of\/} \calm\ {\em by} \hM, and \hM\ will be called the {\em
enveloping
manifold}.

\begin{definition} {\em A {\em boundary point} $p$ of an envelopment
$(\calm,\hM,\phi)$
is a point in the topological boundary of $\phi(\calm)$, i.e.\ a point $p$
belonging to
$\partial(\phi(\calm))= \overline{\phi(\calm)}-\phi(\calm)$ where
$\overline{\phi(\calm)}$ is the closure of $\phi(\calm)$ in \hM.}
\end{definition} The
characteristic feature of such a boundary point is that every open
neighbourhood of it (in
\hM) has non-empty intersection  with $\phi(\calm)$.

\begin{definition} {\em A {\em boundary set} $B$ is a non-empty set of such
boundary  points  (for a fixed envelopment), i.e.\ a non-empty subset of
$\partial(\phi(\calm))$.} \end{definition}

\begin{definition} {\em We shall say that a parametrised curve $\gamma:I\ra
\calm$
{\em approaches} the boundary set $B$
if the curve $\phi\circ\gamma$ has a limit point lying in $B$.}
\end{definition}

\begin{figure} \epsfxsize\hsize
 \leavevmode\centering\epsfbox{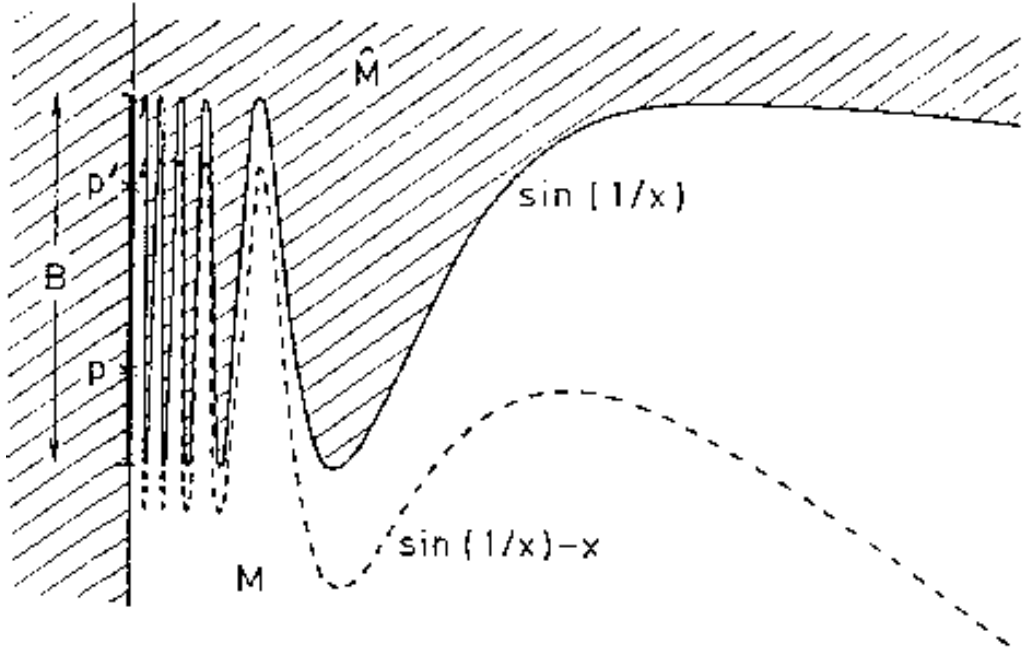}
        \caption{\label{sin1/x} The manifold \calm\ lies below the curve
$y=\sin(1/x)$ and to the right of
the $y$-axis. Points of the boundary set $B$ are not the endpoints of any
curves on \calm.}
\end{figure}

\begin{example} \label{ex:1} {\em It is quite possible to have a boundary point
which is not the endpoint of any curve in \calm. For instance,
let \calm\ be the open submanifold of ${\Bbb R}^{2}$ defined by
$\{(x,y); \,y<\sin(1/x),\; x>0\}$ and let $B$ be the boundary set
$\{(0,y);\,-1<y\leq 1\}$.  All points of $B$ are
limit points of the curve $y=\sin(1/x)-x,\; x>0$, but none of these points is
the endpoint of any curve on \calm\ (see Figure~\ref{sin1/x}).  }
 \end{example}
\begin{definition} {\em If $B'$ is a boundary
set of a
second envelopment $(\calm,\hM',\phi')$ of \calm\ then we say $B$ {\em
covers} $B'$
if for every open neighbourhood \calu\ of $B$ in \hM\ there exists an open
neighbourhood $\calu'$ of $B'$ in $\hM'$  such that \begin{equation}
\phi\circ\phi'^{-1}
\left(\rule[-1ex]{0ex}{3ex}\,\calu' \cap\, \phi'({\calm})\,\right)\, \subset
\,\calu\,. \end{equation}} \end{definition}

\begin{figure}
	\epsfxsize\hsize
	\leavevmode\centering\epsfbox{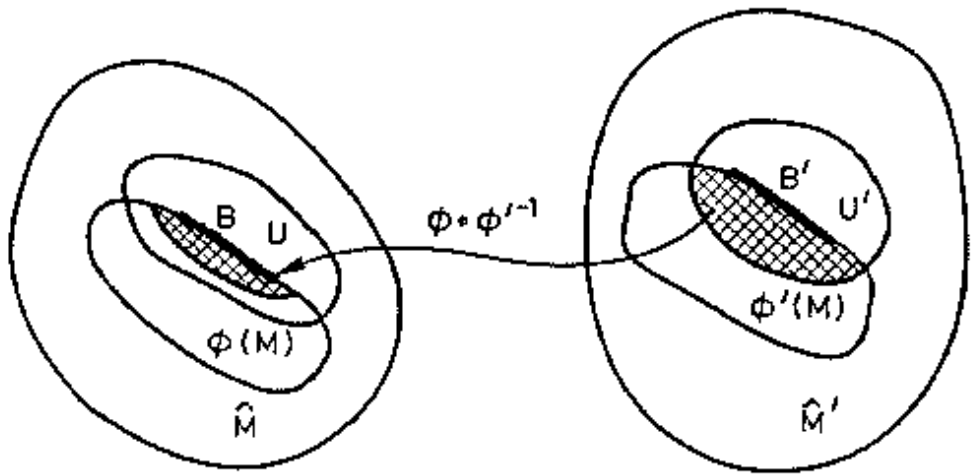}
        \caption{\label{fig:1} The set $B$ covers the set $B'$.}
\end{figure}

The situation is as depicted in Figure~\ref{fig:1}. In effect condition (1)
says that one
cannot get close to points of $B'$ by a sequence of points from within
\calm\ without at the
same time approaching some point of $B$   (see Theorem~\ref{th:crit} for a
more precise
statement of this).

If $p'$ is a boundary point of the envelopment $(\calm,\hM',\phi')$ we shall
say the boundary set $B$ covers (respectively\ is
covered by) $p'$ to mean $B$ covers (is covered by) the singleton boundary
set $\{p'\}$.  Clearly if $p$ is a boundary point lying {\em in} the boundary
set $B$ then $B$ covers $p$. It is also possible, however, for a single
boundary
point to cover a much larger boundary set, as the following example
demonstrates.

\begin{example} \label{ex:2} {\em Let
$\calm = \Rdot{n} = {\Bbb R}^{n}-\{O\}$
and $\phi= \mbox{id}:\calm\ra
{\Bbb R}^{n}$ be the trivial envelopment. Let  $\phi':\calm\ra
{\Bbb R}^{n}$  be
defined by
 \[ \phi'(x_{1},\ldots,x_{n})=\frac{(r+1)}{r}\:(x_{1},\ldots,x_{n}),\mbox{
where }
r=\sqrt{{\textstyle \sum_{i=1}^{n}}x_{i}^{2}}\:.\] The origin $O$ is a
boundary point of $\phi$
and covers the entire boundary set  $S^{n-1}(O,1)$  [the unit sphere centre
$O$] of $\phi'$. }
\end{example}

\begin{theorem} \label{th:coverpt}
A boundary set $B$ covers a boundary set $B'$ if and only if it covers every
boundary point $p' \in B'$.
\end{theorem}
{\em Proof: } The ``only if'' direction is trivial. To prove the ``if''
direction suppose $B$ covers every point $p'\in B'$. Let \calu\ be any open
neighbourhood of $B$ in \hM.
For each $p'\in B'$ let $\calu'_{p'}$ be an open neighbourhood of $p'$ in
$\hM'$ such that
$\phi\,\circ\phi'^{-1}(\calu'_{p'}\cap \phi'(\calm))\subset \calu$. The set
$\calu'= \bigcup_{p'\in B'} \calu'_{p'}$
is clearly an open neighbourhood of $B'$ in $\hM'$ satisfying~(1), hence $B$
covers $B'$.
\qed \vsp

It is commonly of great interest to compare the approach to two boundary sets
along curves in the manifold \calm. In this regard the following theorem is
very useful.
\begin{theorem} \label{th:cover}
If a boundary set $B$ covers a boundary set $B'$ then every curve
$\gamma$ in \calm\ which approaches $B'$ also approaches $B$.
\end{theorem}
{\em Proof:} Suppose $B$ covers $B'$.  Let $\gamma:[a,b)\ra \calm$ be a
curve which
approaches $B'$ and let $p'\in B'$ be a limit point of $\phi'\circ\gamma$.
Suppose $\gamma$ does not approach $B$. Let $t_i \ra b$ be an increasing
infinite sequence of
real numbers such that $\phi' \circ \gamma(t_i) \ra p'$, and set $A=\{\phi
\circ \gamma(t_i); i
\in {\Bbb N} \}$. Then $\calu = \hM - \bar{A} $ is an open
neighbourhood of $B$ and
since  $B$ covers $B'$, there exists an open neighbourhood \,$\calu'$ of
$B'$ in $\hM'$
satisfying condition~(1). As \,$\calu'$ is also an open neighbourhood of
$p'$ there
exists an $n$ such that $\phi' \circ \gamma(t_i) \in \calu'$ for all $i>n$.
Clearly
$\phi\, \circ\, \gamma(t_i) = \phi\, \circ\, \phi'^{-1} \,\circ\, \phi'\,
\circ\, \gamma(t_i)
\in \calu$ for $i>n$, which contradicts $A\, \cap\: \calu\, = \,\mbox{\O}\,$.
Hence $\gamma$ must have a limit point in $B$.
\qed \vsp

It is worth pointing out that the converse to this theorem definitely does not
hold.  For instance, in Example~\ref{ex:1} let $p=(0,-1/2)$ and
$p'=(0,1/2)$. Any curve in \calm\ approaching $p'$ must also approach $p$.
$\{p\}$ does not cover $\{p'\}$, however, since these two points clearly have
non-intersecting open neighbourhoods \calu\ and $\calu'$ and as $\phi =
\phi' =\mbox{id}\:$ in
this example, condition~(1) reduces to $\calu'\cap \calm\subset \calu$ (see
Figure~\ref{sin1/x}).  The following is probably the best that can be said
in this respect.
\begin{theorem} \label{th:cover.conv}
If every curve in \calm\ which approaches a boundary set $B'$ also
approaches a boundary
set $B$, and if every neighbourhood of $B$ in \hM\ contains an open
neighbourhood
\,\calu\ of $B$ whose complement in $\phi(\calm)$ is connected, then $B$
covers $B'$.
\end{theorem}
{\em Proof}: Suppose that all curves $\gamma$ in \calm\ which approach $B'$
also approach $B$, but assume that $B$ does not cover $B'$.
Then by Theorem~\ref{th:coverpt} there exists a $p'\in B'$ such that $B$ does
not cover $p'$.
Hence there exists an open neighbourhood
\calu\ of $B$ in \hM\ such that for every open neighbourhood $\calu'$ of
$p'$ in $\hM'$ the set
$\phi\circ \phi'^{-1}(\calu'\cap\phi'(\calm))$ contains points not
belonging to \calu.
By our hypothesis there is no loss of generality in assuming
$\phi(\calm)-\calu$ to be
a connected set.
Now by paracompactness we can always make the manifold $\hM'$  into
a metric space (e.g.\ by imposing a Riemannian metric on
$\hM'$ and defining $d(x,y)$ to be the shortest distance for
all curves connecting $x$ and $y$).
Let $\calu'_{n} =  \{q'\in \hM'; d(p',q')<1/n\}$. For each $n$ select a point
$p_n \in \phi\circ\phi'^{-1} (\calu'_{n}\cap\phi'(\calm))$ such that it
does not lie
in \calu.  Let $\gamma$ be a curve connecting $p_1$ to $p_2$ to $p_3$ to
\ldots\
and lying entirely in $\phi(\calm)-\calu$. This curve can clearly be made
$C^1$ and does
not have any limit points in $B$. It certainly does have $p'$ as a limit
point, however, since
$\phi'\circ\phi^{-1}(p_n)$ is a sequence of points approaching $p'$.
We therefore have a contradiction and $B$ must cover $B'$.
\qed \vsp

Whilst Theorems~\ref{th:cover} and~\ref{th:cover.conv} represent the best we
can achieve in terms of approaches along parametrised curves in \calm, a
simpler
result holds if one only requires approaches by sequences of points in
\calm. The following
theorem is proved by methods essentially following those of
Theorems~\ref{th:cover}
and~\ref {th:cover.conv}. We therefore omit the proof.

\begin{theorem} \label{th:crit}
$B$ covers $B'$ if and only if for every sequence  $p_1, p_2,\ldots$ of points
in \calm\ such that the sequence $\phi'(p_i)$ has a limit point in $B'$,
the sequence
$\phi(p_i)$ has a limit point in $B$.
\end{theorem}

Covering is a weak partial order on boundary sets:
\begin{enumerate}
        \item $B$ covers $B$.
        \item If $B$ covers $B'$ and $B'$ covers $B''$ then $B$ covers $B''$.
\end{enumerate}
\begin{definition} {\em We say boundary sets $B$ and $B'$ are {\em
equivalent\/} if $B$ covers
$B'$ and $B'$ covers $B$.} \end{definition} This is clearly an equivalence
relation on the set of all
boundary sets.

\begin{definition} {\em An {\em abstract boundary set\/} is an
equivalence class of boundary sets, denoted $[B]$.} \end{definition}

The covering relation passes to abstract boundary sets in the natural way;
we say $[B]$ {\em
covers} $[B']$  if and only if $B$ covers $B'$. This relation is clearly
independent of the choice of representatives. For abstract boundary sets,
however,
the covering relation is a true partial order as it satisfies the further
antisymmetric condition: \[ \mbox{3. If } [B] \mbox{ covers }[B'] \mbox{
and } [B']
\mbox{ covers } [B] \mbox{ then } [B]=[B']. \]

One might be tempted at this stage to define ``abstract boundary points'' as
minimal abstract boundary sets, but any such attempt using an argument
based on Zorn's
lemma is doomed to failure.  Any single boundary point $p$ can always
be blown up to a much larger boundary set in a similar way to
Example~\ref{ex:2}. This is done
by defining a new envelopment with the property that different curves in \calm\
which originally all had $p$ as their endpoints now approach separate endpoints
(e.g., see \cite{Ger}).
Despite this problem we shall make the following definition of an abstract
boundary point.
\begin{definition}
{\em {\bf (The Abstract Boundary)} For a manifold \calm, an
abstract boundary set is an {\em abstract boundary point\/} whenever it has
a singleton
$\{p\}$ as a representative boundary set. In this case the equivalence
class is denoted
by $[p]$. The set of all abstract boundary points will be denoted \bm\ and
called the {\em
abstract boundary} or {\em a-boundary} of \calm.}
\end{definition}

It is to be stressed here that the abstract boundary is defined for {\em every}
manifold \calm, irrespective of the existence of further structure on
the manifold such as a metric, affine connection or chosen family \calc\ of
curves.

It must be realised, however, that an abstract boundary point $[p]$
is no more ``pointlike'' than a more general abstract boundary set $[B]$.
The equivalence class $[p]$ need not even consist only of connected boundary
sets.  For example if one embeds the open interval $\calm=(0,1)$ in the
natural way in the real
line and then embeds it in the unit circle (using angular coordinate $\theta$)
with the map $\theta = \phi(t) = 2\pi t$, then clearly the single boundary
point
$\theta=0$ of the envelopment $(\calm,S^1,\phi)$ is equivalent to the
disconnected
 boundary set $\{t=0,t=1\}$ of the first embedding.

It is not hard to show that every boundary set belonging to the equivalence
class of an abstract boundary point $[p]$ must be a compact set, but whether
this is a necessary and sufficient condition is difficult to resolve.

For the remainder of this section it will be assumed that a family \calc\
of curves
with the b.p.p.\ has been chosen for the manifold \calm\ and this will be
denoted by
(\calm,\,\calc).

\begin{definition} {\em If $(\calm,\hM,\phi)$ is an envelopment of
(\calm,\,\calc) by
\hM, then we say a boundary point $p$ of this envelopment is a {\em
\calc-boundary
point\/} or {\em approachable} if it is a limit point of some curve from
the family
\calc\ (in other words, if some curve of the family approaches $p$).
Boundary points
which are not \calc-boundary points will be called {\em unapproachable}.}
\end{definition}

If a boundary point $p$ covers another boundary point
$p'$ and $p'$ is a \calc-boundary point, then it is clear by
Theorem~\ref{th:cover} that so is $p$, since any
curve which approaches $p'$ must also approach $p$. This leads naturally to the
following definition. \begin{definition}\label{def:21} {\em An abstract
boundary point
$[p]$ is
an {\em abstract \calc-boundary point\/}, or simply {\em approachable},
if $p$ is a \calc-boundary point. Similarly, an abstract boundary point $[p]$
is
{\em unapproachable} if $p$ is {\em not} a \calc-boundary point.}
\end{definition}

This definition is clearly independent of the choice of representative
boundary point
$p$. We denote the set of abstract \calc-boundary points of \calm\ by
${\cal B}_{\cal
C}(\calm)$. Note how properties such as being approachable only have to be
preserved
one way under covering in order to pass to the abstract boundary. For
example, the
converse of the statement just prior to Definition~\ref{def:21} is certainly
{\em not} true---\cbps\ {\em may} cover unapproachable boundary points, as the
following example demonstrates.

\begin{example} \label{ex:3} {\em Consider the manifold
$\calm=\{(x,y) \in {\Bbb R}^{2};\;-\infty<x<\infty,\; 0<y<\infty\}$
with the metric d$s^{2}=y^{2}\,$d$x^{2}\:+\:$d$y^{2}$ (this is the
covering space of the cone \cite{ES}, or the Riemann surface of $\log(z)$).
We set $\calc = \calc_g (\calm)$, the family of affinely parametrised geodesics
on \calm.
Let $\phi: \calm \ra {\Bbb R}^{2}$ be the trivial envelopment defined by
$\phi\,(x,y)=(x,y)$. Every boundary point on $y=0$ is a
\cbp, since it is the endpoint of a vertical geodesic
$x=\mbox{constant}$. Furthermore, these are the {\em only} geodesics which
approach the boundary $y=0$, since the general geodesic is given by
$y=\alpha \sec(x-x_{0}), \alpha >0,\, |x-x_{0}|<\pi/2$\,.
A second envelopment $\phi':\calm \ra{\Bbb R}^{2}$ defined by
\[(x',y')=\phi'(x,y)=
\left(\rule[-0.5ex]{0ex}{4ex}\,\frac{x}{y}\:,\sqrt{x^{2}+y^{2}}\;\right)\]
again maps
\calm\ onto the region $y'>0$ of \,${\Bbb R}^{2}$. It has the effect
of spreading
out the vertical geodesics $x=a\neq 0$ so that they approach $y'=|a|$ as $x'\ra
\pm\infty$. Thus  on the boundary $y'=0$ only the origin $(0,0)$ is a
\calc-boundary
point, all others $(x'\neq 0\,,y'=0)$ being  unapproachable.  It is also
easily seen that
all these unapproachable boundary  points are covered by the \cbp\ $(0,0)$
of the
original envelopment $\phi$.} \end{example}

\section{Pseudo-Riemannian manifolds}
\subsection*{Extensions} \vsp
We now introduce a metric on \calm. In order to establish conventions, we
first give a standard
definition.

\begin{definition} {\em A $C^{k}$ {\em metric\/} $g$ on \calm\ is a second
rank covariant
symmetric and non-degenerate $C^{k}$ tensor field on \calm. The pair
$(\calm,g)$ is called a
$C^{k}$ {\em pseudo-Riemannian manifold}. When $g$ is positive definite
it is called {\em Riemannian}.} \end{definition} Our discussion will refer
to metrics of any
signature unless specifically stated otherwise. An envelopment of a
pseudo-Riemannian manifold
will be denoted $(\calm,g,\hM,\phi)$. The metric ${({\phi^{-1}})^{*}}g$
induced on the open
submanifold $\phi(\calm)$ of \hM\ by the $C^{\infty}$ embedding $\phi$ will
also be denoted by
$g$ when there is no risk of ambiguity.

\begin{definition} {\em A $C^{l}$ {\em extension} $(1\leq l\leq k)$ of a
$C^{k}$
pseudo-Riemannian manifold $(\calm,g)$ is an envelopment of it by a $C^{l}$
pseudo-Riemannian
manifold $(\hM,\hat{g})$ such that
\[ \hat{g}|_{\phi({\cal M})} = g, \]
denoted $(\calm,g,\hM,\hat{g},\phi)$.
When $l=k$, we talk simply of $(\hM,\hat{g})$ being an
{\em extension\/} of $(\calm,g)$.} \end{definition}

We note that this definition of an extension of a pseudo-Riemannian
manifold $(\calm,g)$ can
be applied in a precisely analogous manner to a manifold \calm\ simply
endowed with a $C^{k}$
affine connection $\nabla$, denoted $(\calm,\nabla)$. Furthermore, this
will also be true
for all definitions to follow. It is important to keep this in mind, since
it means that our
scheme can be applied to a wide class of theories including conformal,
projective and
gauge theories.

\subsection*{Regular Boundary Points} \vsp \begin{definition} {\em We say a
boundary
point $p$ of an envelopment $(\calm,g,\hM,\phi)$ is  $C^{l}$ {\em regular
for} $g$ if there
exists a $C^{l}$ pseudo-Riemannian manifold $(\bM\,,\bar{g})$ such that
$\phi(\calm)\cup\{p\}
\subseteq \bM \subseteq \hM$ and $(\calm,g,\bM\,,\bar{g},\phi)$ is a
$C^{l}$ extension of
$(\calm,g)$.} \end{definition} Note that we require the {\em same\/}
mapping $\phi$ for the
extension as for the original envelopment (although strictly speaking,
since the target set
\bM\ is different to the target set \hM\ of $\phi$, it should be given a
new name $\bar{\phi}$
defined by the requirement that $\bar{\phi}(q)=\phi(q)$ for all $q \in \calm$).

A $C^{k}$ regular boundary point will simply be called {\em regular}.
Although there is no serious loss of generality in using this term (since if
$l<k$ we may simply regard
$(\calm,g)$ as being a $C^{l}$ pseudo-Riemannian manifold), we shall
persist in our terminology because the distinction does become important for
singular boundary points.

The notion of regularity cannot, however, be transferred to
the abstract boundary as it stands, since it is not invariant under
equivalence of
boundary points. The following simple example clearly demonstrates this fact.

\begin{example} \label{ex:4}
{\em Embed the one-dimensional manifold $\calm=(0,1)$ with metric d$s^{2}
=\;$d$x^{2}$ into the manifold $\Bbb R$ in two ways: $y=\phi(x)=x$
and $z=\phi'(x)
=x^{1/2}$. The boundary points $y=0$ and $z=0$ are clearly equivalent by our
earlier definitions, but the first is $C^{\infty}$ regular while the second is
not $C^l$ regular for any $l\geq 1$. This follows because the metric induced
by the second embedding,
\[{\rm d}s^{2}= \,4 z^{2} \,{\rm d}z^{2},\]
is degenerate at $z=0$ and cannot therefore be extended to any open
interval $(-a,1)$, where $a>0$. Thus the abstract boundary point in
question has at least two
representative  boundary points, one of which is regular while the other is
not.}
\end{example}

The notions of an {\em extension} of a pseudo-Riemannian manifold
$(\calm,g)$ and a
{\em regular boundary point} of an envelopment $(\calm,g,\hM,\phi)$ are both
completely independent of whether or not a family \calc\ of curves with the
b.p.p.\
has been chosen for \calm. This will not be true for the notions which
follow, such as
a ``point at infinity" and a ``singular boundary point". Therefore we shall
henceforth always
assume that our pseudo-Riemannian manifold is endowed with a family \calc\
of curves with
the b.p.p.\ which normally (i.e.\ unless otherwise specified) includes the
family of all
geodesics with affine parameter $\calc_{g}(\calm)$.  The general situation
will be denoted
$(\calm,g,\calc)$ while the usual notation $(\calm,g)$ will be reserved for
the case where
$\calc = \calc_{g}(\calm)$. An envelopment of a pseudo-Riemannian manifold
with a family of
curves satisfying the b.p.p.\ will be denoted $(\calm,g,\calc,\hM,\phi)$.

The following example shows that regular boundary points can even be
unapproachable by
geodesics in \calm.   \begin{example} \label{ex:5}  {\em  Let \calm\ be the
open submanifold of
\,${\Bbb R}^{2}$ defined by $y>x^{1/2}, x>0$ and let $g$ be the usual
flat metric
d$s^2=\;$d$x^2\:+\:$d$y^2$. The boundary point $(0,0)$ is $C^{\infty}$
regular since the metric
extends to all of ${\Bbb R}^{2}$, yet it is clearly unapproachable by
any geodesics (straight
lines) in \calm.} \end{example}

\subsection*{Points at Infinity}
 \vsp The non-regular boundary points can be broken up into two groups,
those which
are \cbps, also called approachable, and the rest which are unapproachable.
Although we have seen examples of regular boundary points which are
unapproachable
(Example~\ref{ex:5}), the non-regular unapproachable boundary points do not
seem to
merit serious further discussion.  This is because they usually occur when
we blow up a region of a boundary by spreading out a family of approaching
curves
too thinly (e.g.\ Example~\ref{ex:3}).

The approachable non-regular boundary points do, however, have a very rich
structure.
First one must ask of them whether one can effectively ever ``get there''
along a
curve in \calc\ with a finite value of the parameter or not. To this end we
introduce the
concept of a point at infinity.

\begin{definition}\label{pai def} {\em Given an $(\calm,g,\calc)$ we will say
that a
boundary point $p$ of
the envelopment $(\calm,g,\calc,\hM,\phi)$ is a $C^l$ {\em point at
infinity for \calc\/} if
\begin{quote} (i) $\;\;p$ is not a $C^l$ regular boundary point, \\
(ii) $\:p$ is a \cbp, and \\
(iii) no curve of \calc\ approaches $p$ with bounded parameter.
\end{quote}} \end{definition}

Condition (iii) says that for no interval $I=[a,b<\infty)$ is
there a curve $\gamma:I\ra \calm$ in the family \calc\ and an increasing
infinite sequence of
real numbers $\{t_{i}\}$ in $I$ such that
\[\phi\,(\gamma(t_{i}))\lra p\: \mbox{ as }\: t_{i}\ra b\,.\]
Clearly a $C^l$ point at infinity is also a $C^{l'}$ point at infinity
for all $l'>l$. In particular, it is always a $C^k$ point at infinity and there
is no real loss of generality in simply calling it a {\em point at infinity}.

Note that by the bounded parameter property, the concept of a point at infinity
is independent of the choice of parametrisation on the curves from \calc\
which approach $p$. It is here that the importance of imposing the bounded
parameter property on \calc\ becomes evident.

Condition~(i) ensures that no boundary point is classified as both regular and
a point at infinity. Without it, such boundary points do, in fact, occur as
made clear by the following two examples.
\begin{example} \label{ex:6} {\em In Example~\ref{ex:5}
let \calc\ consist of all the geodesics in \calm\ supplemented with the curves
$x(t)=1/t^2$, $y(t)=C/t$, $C>1,\,1\leq t<\infty$. The boundary point $(0,0)$ is
still regular (nothing has changed as regards the metric) but it is ``at
infinity'' for \calc,
since it is approachable only by curves with unbounded parameter.}
\end{example}

This example is somewhat artificial, in that the added curves, and more
particularly the choice of their parametrisation, seem to have nothing to do
with the metric. A rather more subtle example is the following, in which
there is a regular
boundary point which is geodesically approachable, but only by geodesics with
unbounded affine parameter.

\begin{figure}
	\leavevmode\centering\epsfbox{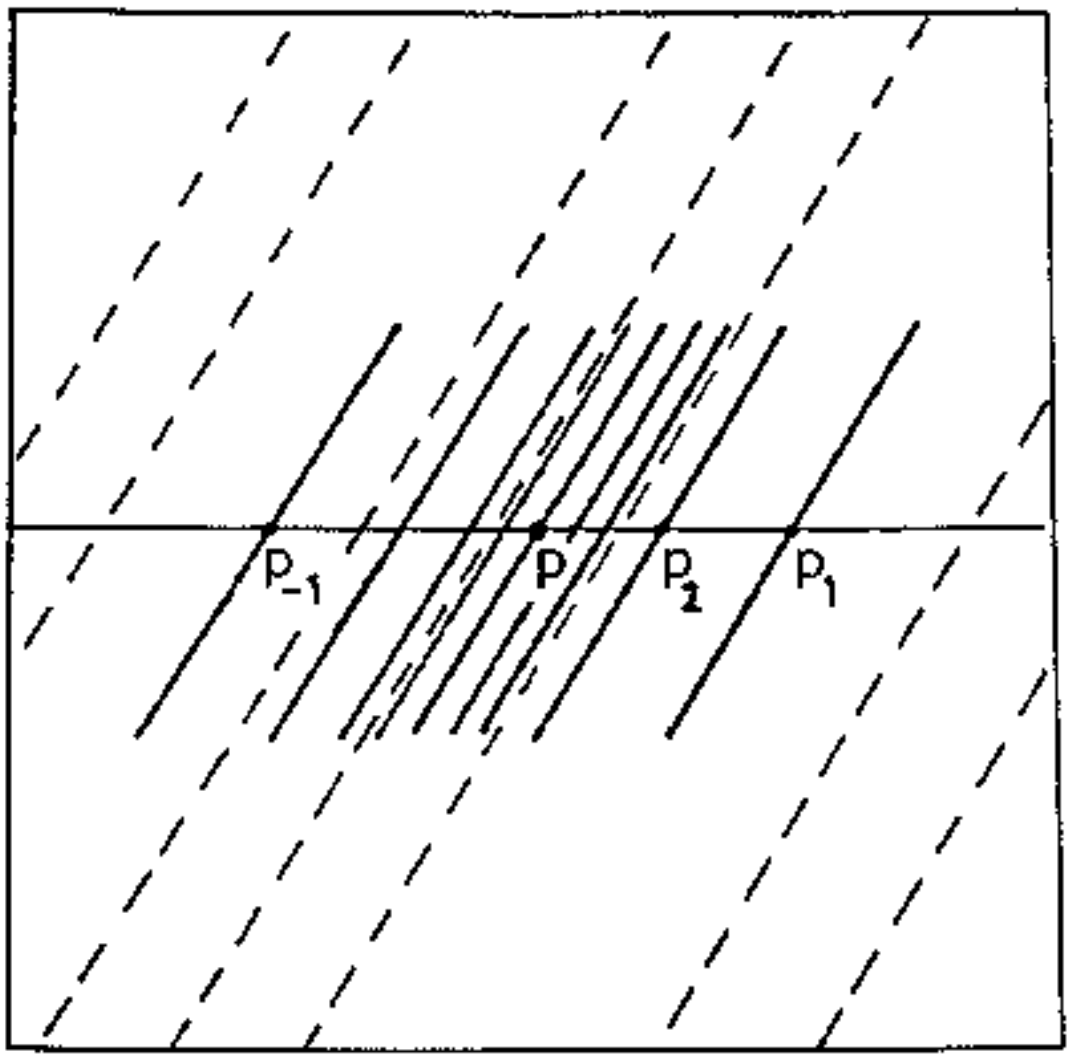}
        \caption{\label{fig:torus} An example of a $C^\infty$ regular
boundary point which is
approachable only by geodesics in \calm\ with infinite affine parameter.}
\end{figure}

\begin{example} \label{ex:torus}
{\em Let \hM\ be the unit torus, i.e.\ ${\Bbb R}^{2}/{\Bbb Z}^{2}$, with the
usual
flat metric d$s^2=\;$d$x^2\:+\:$d$y^2$. Let $\gamma$ be the geodesic in
\hM\ generated by the line
$x=t\,,\,y=t\sqrt{2}\: (t\geq 0)$ and let $p$ be the point
$(\frac{1}{2},\frac{1}{2})$.
On the central line $L=\{(x,1/2)\,;\,0\leq x<1\}$ choose points
\[p_{\pm
i}\,=\left(\rule[-0.5ex]{0ex}{4ex}\,\frac{1}{2}\left(\rule[-0.5ex]{0ex}{4ex}
1 \pm
\frac{1}{2^i}\,\right),\,\frac{1}{2}\,\right), \hspace{5em}i=1,2,3,\ldots
\] For each $i=\pm
1,\pm 2,\ldots$ let $L_i$ be the closed line segment of  length 1/2 and
slope $\sqrt{2}$
centred on the point $p_i$ and let $L_0$ be a similar line segment with
centre $p$. Now define
\calm\ as the open submanifold of \hM\ consisting of the  complement in
\hM\ of this infinite
set of closed line segments,
\[\calm=\hM - \bigcup_{i \in \Bbb Z} L_i\]
(see Figure~\ref{fig:torus}). Clearly $p$ is a boundary point of the
envelopment $(\calm,\hM,
\mbox{id}_{\cal M})$ and is $C^{\infty}$ regular. Now apart from
its starting
point at $(0,0)$, the geodesic $\gamma$ does not pass through any point
$(x,y)$ where $x$ and
$y$ are both rational and, in particular, it does not pass through any of
the points $p_i$ or
$p$ on $L$. It follows that $\gamma$ lies completely in \calm. Furthermore,
there is an
increasing infinite sequence of positive numbers $t_n \ra \infty$ such that
$\gamma(t_n)$ all lie on $\calm\,\cap\, L$ and such that $\gamma(t_n)\ra
p$. Thus $\gamma$
approaches $p$ and does so with unbounded affine parameter. The same is true
of every geodesic in \hM\ with slope $\sqrt{2}$ which does not pass
through any of the points $p_i$ or $p$.  Moreover these are the {\em only}
geodesics in \calm\ which approach $p$. Thus $p$ is $C^\infty$ regular yet it
is like a point at infinity with respect to geodesics in \calm.}
\end{example}

The key thing about these two examples is that regularity always takes
precedence over being ``at infinity''.  Thus even if one can only ``reach" the
boundary point along curves from \calc\ having unbounded parameter, if it is
regular then it will {\em not} be classified as a
point at infinity. Suppose, however, that in Example~\ref{ex:6} one were to
perform a ``bad'' coordinate transformation, e.g., $x=x'^{3}, y=y'^{3}$.
If the manifold \calm\ is re-embedded in ${\Bbb R}^{2}$ with the new
coordinates $x', y'$ taken as rectangular, then the
origin definitely does become a point at infinity, since conditions~(i), (ii)
and~(iii) of Definition~\ref{pai def} are now all satisfied. This point at
infinity
is, however, in
a sense ``artificial'' or ``removable'', since it is covered by (indeed
equivalent to) the original regular boundary point at the origin.
A similar bad transformation can be applied in a neighbourhood of the
boundary point
$(\frac{1}{2},\frac{1}{2})$ of Example~\ref{ex:torus},
also converting it into a point at infinity.

\begin{definition} {\em We define a boundary point which is a point at
infinity to be {\em
removable\/} if it can be covered by a boundary set of another embedding
consisting entirely of
regular boundary points. When a point at infinity is not removable, it will
be called {\em
essential}.} \end{definition} In a sense essential points at infinity are
boundary points
which really do have a component at infinity (i.e., which cannot be
transformed away).
Furthermore the concept of being an essential point at infinity passes to
the abstract
boundary, as the following theorem demonstrates.
\begin{theorem} \label{th:infty}
If the boundary point $p$ is an essential point at infinity and $p$ is
equivalent to the
boundary point $p'$, then $p'$ is also an essential point at infinity.
\end{theorem}
{\em Proof:} Since $p'$ covers $p$, it follows from Theorem~\ref{th:cover} that
it must
be a \cbp\ (since
$p$ is a \cbp). $p'$ cannot be regular, else $p$ would be covered by a
regular boundary
point, contradicting it being an essential point at infinity. Now since
$p$ also covers $p'$, $p'$ cannot be the limit point of any curve in \calc\
with
bounded parameter else, by Theorem~\ref{th:cover}, $p$ would not be a point
at infinity.  Hence $p'$ is a point at infinity. Furthermore it is an essential
point at infinity for if $B$ is any boundary set of regular boundary points
which covers $p'$, then by transitivity of the covering relation this
boundary set would also cover $p$, again contradicting its essentialness.
\qed \vsp

It is interesting to note that an essential point at infinity may itself cover
regular boundary points. For example, if one is given an envelopment of a
manifold which has
both an unapproachable regular boundary point and an essential point at
infinity, then it is in
general a straightforward matter to create a new envelopment in which these two
points are ``joined together'' to coalesce into a single new boundary point.
This newly created boundary point would also be an
essential point at infinity, but would cover the original regular boundary
point.

\begin{definition} {\em An essential point at infinity which covers a
regular boundary point
will be called a {\em mixed point at infinity}. Otherwise it will be termed
a {\em pure point
at infinity}.} \end{definition} It is easy to see that both these
categories are invariant under boundary point
equivalence and therefore pass to the abstract boundary.

\subsection*{Singular Boundary Points} \vsp \begin{definition}\label{sing
def}{\em A
boundary point $p$ of an
envelopment $(\calm,g,\calc,\hM,\phi)$ will be called $C^{l}$ {\em
singular\/} or a $C^{l}$
{\em singularity\/} if  \begin{quote}
(i) $\;\;p$ is not a $C^{l}$ regular boundary point,\\
(ii) $\:p$ is a \cbp, and \\
(iii) there exists a curve from \calc\ which approaches $p$ with bounded
parameter.
\end{quote}
Alternatively, one could say $p$ is $C^l$ singular if it is a \cbp\ which is
not $C^l$ regular and not a $C^l$ point at infinity.} \end{definition}

Since a boundary point which is not $C^{l}$ regular is clearly not $C^{l'}$
regular
for all $l'\geq l$ it follows at once that if $p$ is $C^{l}$ singular then
it is
$C^{l'}$ singular for all $l'\geq l$---in particular, it is always $C^{k}$
singular. In general we shall simply say that $p$ is {\em singular\/} if it
is $C^{k}$ singular (i.e., if $p$ is $C^{l}$ singular for some $l \leq k$).

\begin{example} \label{ex:clsing}
{\em Consider the metric
\[{\rm d}s^2 =(1+r^{2n}){\rm d}r^2 + r^2 {\rm d}\theta^2, \mbox{ where }
r>0,\; 0 \leq \theta < 2\pi\]
on the manifold $\calm= \,\Rdot{2}$ (see Example~\ref{ex:2}).
This is the metric induced on the surface in ${\Bbb E}^{3}$
obtained by rotating the curve $z=r^{n+1}/(n+1)$ about the $z$-axis.
The curvature scalar is readily shown to be proportional to
$r^{2n-2}/(1+r^{2n})^2$ which $\ra \infty$ as $r\ra 0$  for $0<n<1$.
In fact for $\frac{1}{2}<n<1$ one finds that the boundary point $r=0$ is
$C^1$ regular but
$C^2$ singular. Similarly for $N<2n<N+1$, where $n\in{\Bbb N}$, we
have that $r=0$ is
$C^N$ regular but $C^{N+1}$ singular. These results are easily seen by
considering \calm\ as being embedded in ${\Bbb R}^{2}$ with the
usual polar
interpretation of the coordinates $(r,\theta)$ and then transforming to
rectangular coordinates
$(x,y)$.} \end{example}

\begin{example} \label{ex:rplus1}
{\em Consider the metric
\[{\rm d}s^2= {\rm d}r^2+(r+1)^2 {\rm d}\theta^2, \mbox{ where }
0<r<\infty,\; 0\leq\theta<2\pi\]
on the manifold $\calm= \,\Rdot{2}$. In this case the boundary point $r=0$ is
singular if we use the natural polar embedding of \calm\ in ${\Bbb R}^{2}$. It
is
worth considering this example in some detail since at first sight the
metric shows no
pathological behaviour at the boundary point in question. The coordinates
$(r,\theta)$ do not,
however, constitute a coordinate patch for the manifold ${\Bbb R}^{2}$ in a
neighbourhood of $r=0$. It is therefore necessary to perform a transformation
to
``rectangular'' coordinates $\,x=r\cos\theta\,,\, y=r\sin\theta\,$,
whereupon the metric becomes
\[{\rm d}s^2 = \,\frac{(r+1)^2}{r^2}\,({\rm d}x^2+{\rm d}y^2)\, -
\,\frac{2r+1}{r^4}\,(x{\rm d}x
+y{\rm d}y)^2\,. \] It is easy to see that $x=y=0$ is not a regular
boundary point since each metric
component becomes infinite for almost all directions of approach. It is
clearly not a point at
infinity since it is approached by geodesics with bounded parameter, hence
it is a singularity.
Let us now re-embed \calm\ in ${\Bbb R}^{2}$ using the $\phi'$ of
Example~\ref{ex:2}.
Then $\phi'(\calm)$ is the region $r>1$ of ${\Bbb R}^{2}$ and again
using polar
coordinates on ${\Bbb R}^{2}$, the induced metric on $\phi'(\calm)$
becomes the
standard flat metric d$s^{2}=\;$d$r^{2}+r^{2}$d$\theta^2$. This can, of
course, be
extended to all of ${\Bbb R}^{2}$ using rectangular coordinates
$(x,y)$. Thus the
original singular boundary point $r=0$ is equivalent to the boundary set
$S^{1}(O,1)$ which is made up entirely of $C^{\infty}$ regular boundary
points.
Such a singularity will be termed ``removable''. We begin to formalise the
message of this
example with the following definition.} \end{example}

\begin{definition} {\em A boundary set $B$ will be called $C^{l}$ {\em
non-singular\/} if none
of its points are $C^{l}$ singular, i.e., if they are all either $C^l$
regular, $C^l$ points at
infinity or unapproachable boundary points. (N.B.\ As discussed above, the
first and last
categories are not mutually exclusive.)} \end{definition}

As in Example~\ref{ex:rplus1} singular boundary points can arise which are
equivalent to $C^{l}$
non-singular boundary sets. Such boundary points should not be classified
as ``truly'' or
``essentially'' singular and will be called ``removable
singularities''. Other examples are the one-dimensional Example~\ref{ex:4}
above and the boundary points $(r=2m\,,t=\mbox{constant})$ of the
Schwarzschild ``singularity'',
which are all covered by the $C^{\infty}$ regular boundary point
$(u=0,v=0)$ of the
Kruskal--Szekeres extension \cite{Krusk,GSzek}. A more precise definition
of this concept is now
given.

\begin{definition} {\em A $C^{m}$ singular boundary point $p$ will be called
$C^{m}$ {\em removable\/} if it can be covered by a $C^{m}$ non-singular
boundary set $B$. Clearly if $p$ is a $C^{m}$ removable singularity, then
for all $m'\leq m$ such
that $p$ is $C^{m'}$ singular, it is also $C^{m'}$ removable. If $m=k$ we say
simply that $p$ is a {\em removable singularity}.} \end{definition}

\begin{definition} {\em A $C^{m}$ singular boundary point $p$ will be
called $C^{m}$ {\em essential} if it
is not $C^{m}$ removable.  If $p$ is $C^{m}$ essential then it is $C^{m'}$
essential for all $m'\geq m$, whence it is always $C^{k}$ essential and
we can describe $p$ simply as an {\em essential singularity}.} \end{definition}

Keeping track of all these orders of differentiability is a tedious business
and from now on we will simply use the terms regular, singular,
removable, essential, etc.\ in most cases. It is usually a straightforward
matter to discover corresponding statements for more general orders of
differentiability than $k$. It is worthwhile keeping in mind the following
easily proved theorem, stating in essence that singularities can never be
removed ``to infinity''.

\begin{theorem} \label{th:removs}
Let $p$ be a removable singularity and let $B$ be any non-singular
boundary set which covers $p$. Then $B$ contains at least one regular
boundary point.
\end{theorem}
{\em Proof:} By Definition~\ref{sing def}, there is a curve $\gamma$ from
\calc\ which
approaches $p$ with bounded parameter. Since $B$ covers $p$, $\gamma$
must also approach $B$ (by Theorem~\ref{th:cover}). Let $q \in B$ be any
limit point of the curve
$\gamma$. $q$ is clearly a \cbp\ and since the curve $\gamma$ approaches it
with bounded parameter,
it is not a point at infinity. Hence $q$ must be a regular boundary point.
\qed \vsp

It is useful to further subclassify essential singularities and to this end
we provide the
following definition.

\begin{definition} {\em An essential singularity $p$ will be called a {\em
mixed\/} or {\em
directional singularity\/} if $p$ covers a boundary point $q$ which is
either regular or a point at
infinity. Otherwise, when $p$ covers no such boundary point, we shall call
it a
{\em pure singularity}.} \end{definition}

These notions seem to encapsulate earlier discussions of ``directional
singularities'' as they
appear in the literature, particularly with regard to the Curzon solution
\cite{Stach,Gaut,Coop}.
Possibly the word ``mixed'' is better to use in this context since such a
singularity might only cover an unapproachable regular boundary point,
which would hardly
make the behaviour dependent on the ``direction of approach''.  Nevertheless
we will persevere with the more standard terminology and usually call such
singularities directional.

\begin{figure}
	\epsfxsize\hsize
	\leavevmode\centering\epsfbox{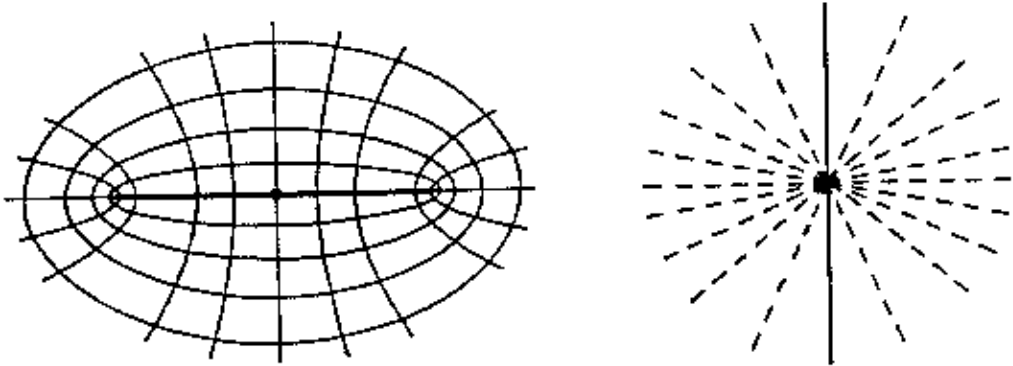}
        \caption{\label{fig:dirsing} A directional singularity. A curvature
singularity exists at the origin of the $(x,y)$ coordinate system on the left.
The origin on the right induced by the use of elliptical coordinates is
a directional singularity, as it covers the whole line segment $[-1,1]$ of the
original $x-$axis, which consists of both a singular boundary point and
infinitely many regular
boundary points. } \end{figure}

\begin{example} \label{ex:dirsing}
{\em Consider the metric
\[{\rm d}s^2=F^2(r)\,({\rm d}r^2 +r^2 {\rm d}\theta^2), \mbox{ where }
r>0,\; 0\leq\theta<2\pi\
\mbox{and } F^2(r)=1+r^n \] defined on the two-dimensional manifold $\calm=
\,\Rdot{2}$. When
$0<n<2$ it is seen that the curvature scalar \[ R_{\mbox{\scriptsize
curv}}= - \:\frac{n^2
r^{n-2}}{(1+r^n)^3}\, \ra -\,\infty \:\mbox{ as }\: r\ra 0\,, \] so that
the boundary point $r=0$ of
the natural embedding of \calm\ in ${\Bbb R}^{2}$ is a $C^{2}$
singularity (it is clearly
not a point at infinity). Now transform to elliptical coordinates $(\eta,
\psi)$ given by \[
\begin{array}{ccccc} x & = & r \cos \theta & = & \cosh \eta \cos \psi\:, \\
y & = & r \sin \theta & = & \sinh \eta \sin \psi\:.
\end{array} \]
In these coordinates the metric becomes
\[{\rm d}s^2 = F^2(r)\,({\rm cosh}^2 \eta - {\rm cos}^2 \psi)\, ({\rm
d}\eta^2 + {\rm d}\psi^2)
\hspace{.2in} (0<\eta<\infty,\; 0\leq\psi<2\pi) \] where
\[r^2 = \cosh^2\!\eta - \sin^2\!\psi\:. \]
In the $x-y$ plane, $\eta=0$ corresponds to the strip $-1\leq x \leq1$ of
the $x$-axis.
If we now perform a transformation to ``rectangular'' coordinates $(X,Y)$ based
on $(\eta,\psi)$ as ``polars'', viz.,
\[ X=\eta \cos\psi\:,\mbox{ } Y= \eta \sin\psi\:, \]
the metric assumes the rather prohibitive form
\[ {\rm d}s^2 = \frac{F^2(r)}{\eta^4}\:\left(\rule[-0.5ex]{0ex}{4ex}
\cosh^2\!\eta - \frac{X^2}{\eta^2}\,\right)\:[\:(\eta^2 X^2 + Y^2)\,{\rm
d}X^2 + (\eta^2 Y^2 +
X^2)\,{\rm d}Y^2\]\[\;\;\;\;\;\;\;+2XY(\eta^2 -1)\,{\rm d}X{\rm d}Y\:]\:,\]
where
\[ r^2 = \cosh^2\!\eta - \frac{Y^2}{\eta^2}\:,\hspace{3em} \eta=\sqrt{X^2 +
Y^2}\:. \]

Now suppose that we are presented with this metric in $(X,Y)$ coordinates, but
without any words of explanation as to its origin. Taking $(X,Y)$ as
rectangular coordinates on
the manifold ${\Bbb R}^{2}$, we wish to classify the boundary
point $(0,0)$ of
$\,\Rdot{2}$. From evaluation of the curvature scalar, which must come to
the same along a given
curve as in the first embedding, one sees that  $R_{\mbox{\scriptsize
curv}}\ra -\,\infty$ as one
approaches the origin $X=Y=0$ ($\eta = 0$) along the $Y-$axis. For any
other direction of
approach $X=\alpha Y$ ($\alpha\neq 0$), however, $R_{\mbox{\scriptsize
curv}}$ can  be shown to have
a finite limit (see Figure~\ref{fig:dirsing}). Typically this is the cue for a
directional singularity. The way our example has been constructed it is a
straightforward matter to
see how the directionality may be unravelled. The boundary point $X=Y=0$ is
equivalent
to the strip $-1\leq x \leq1$ of the $x-$axis $(\eta = 0)$ in the original
embedding, which consists
of both a $C^{2}$ essentially singular boundary point and infinitely many
$C^{2}$ regular
boundary points. It follows that the origin of the second embedding is a
$C^{2}$ directional
singularity.} \end{example}

The following theorem implies that the property of being an essential
singularity passes to the abstract boundary.

\begin{theorem} \label{th:4}
If a boundary point $p$ of an envelopment $(\calm,g,\calc,\hM,\phi)$
covers a boundary point $p'$ of a second envelopment
$(\calm,g,\calc,\hM',\phi')$ which is essentially singular, then $p$ is also an
essential singularity.
\end{theorem}
{\em Proof:} The boundary point $p$ is a \cbp\ since it covers the \cbp\
$p'$. It is neither
regular nor a point at infinity, else $p'$ would be covered by the non-singular
boundary set $\{p\}$ and hence would be removable (i.e.\ not essential).
Furthermore it must be an essential singularity, for if it were covered by a
non-singular boundary set $B$, then $B$ would also cover $p'$ by the
transitivity
of the covering relation, again contradicting the assumption that $p'$
is an essential singularity.
\qed \vsp

\section{Classification of boundary points}
Suppose that we are given an envelopment $(\calm,g,\calc,\hM,\phi)$ of a
$C^k$ pseudo-Riemannian
manifold $(\calm,g,\calc)$ and wish to classify a specific boundary point $p$.
We proceed by asking a series of questions. Each question is in principle
decidable, though we do not mean to imply by this that the decision is always
easy to carry out.  The questions to be decided are: \\[2mm]
(1) {\em Is $p$ a $C^l$ regular boundary point for some $l\leq k$?} This
is usually a fairly straightforward thing to decide. If $p$ is regular then
there
must exist an open coordinate neighbourhood \calu\ of $p$ in \hM\
and a metric which extends $g$ in a $C^l$ manner from its restriction to
$\,\calu\,\cap\,\calm$ to
all of \calu. If the answer is YES, we are finished, except for possibly
enquiring whether
$p$ is approachable (i.e.\ a \cbp) or not.  This latter question is again
answerable by investigating whether $p$ is a limit point of some curve
from \calc. \vsp

For convenience, let us assume from now on that $l=k$
(corresponding questions
to be decided for different orders of differentiability are easily posed).
If the answer to question (1) is NO, then we must proceed as follows:\\[2mm]
(2) {\em Is $p$ a \cbp?} If NO, then $p$ is filed away as an unapproachable
non-regular boundary point. These are essentially uninteresting points, though
a further investigation might be carried out to ascertain whether they cover
any (unapproachable) regular boundary points.
If YES, then we proceed with the following questions:\\[2mm]
(3)  {\em Is there a curve from \calc\ which approaches $p$ with bounded
parameter?} If NO, then $p$ is a point at infinity, while if YES, then
it is a singularity. \\[2mm]
If $p$ is a point at infinity we ask:\\[2mm]
(4) {\em Is $p$ covered by a boundary set $B$ of another embedding
consisting only of regular
and/or unapproachable boundary points?} If YES, then the point at infinity
is called removable,
while if NO, then it is called essential. In the latter case we proceed to
ask:\\[2mm]
(5) {\em Does $p$ cover a regular boundary point $q$ of another embedding?}
If YES, then $p$ is
a mixed point at infinity, while if NO, then it is a pure point at
infinity.\\[2mm]
If $p$ is singular we ask:\\[2mm]
(6) {\em Is $p$ covered by a non-singular boundary set $B$ of another
embedding?} If YES, then it is
a removable singularity, while if NO, then it is an essential singularity
and we
can ask further: \\[2mm]
(7) {\em Does $p$ cover any regular boundary points or points at infinity
of other embeddings?} If
YES, then it is a directional singularity, while if NO, then it is a pure
singularity. \vsp

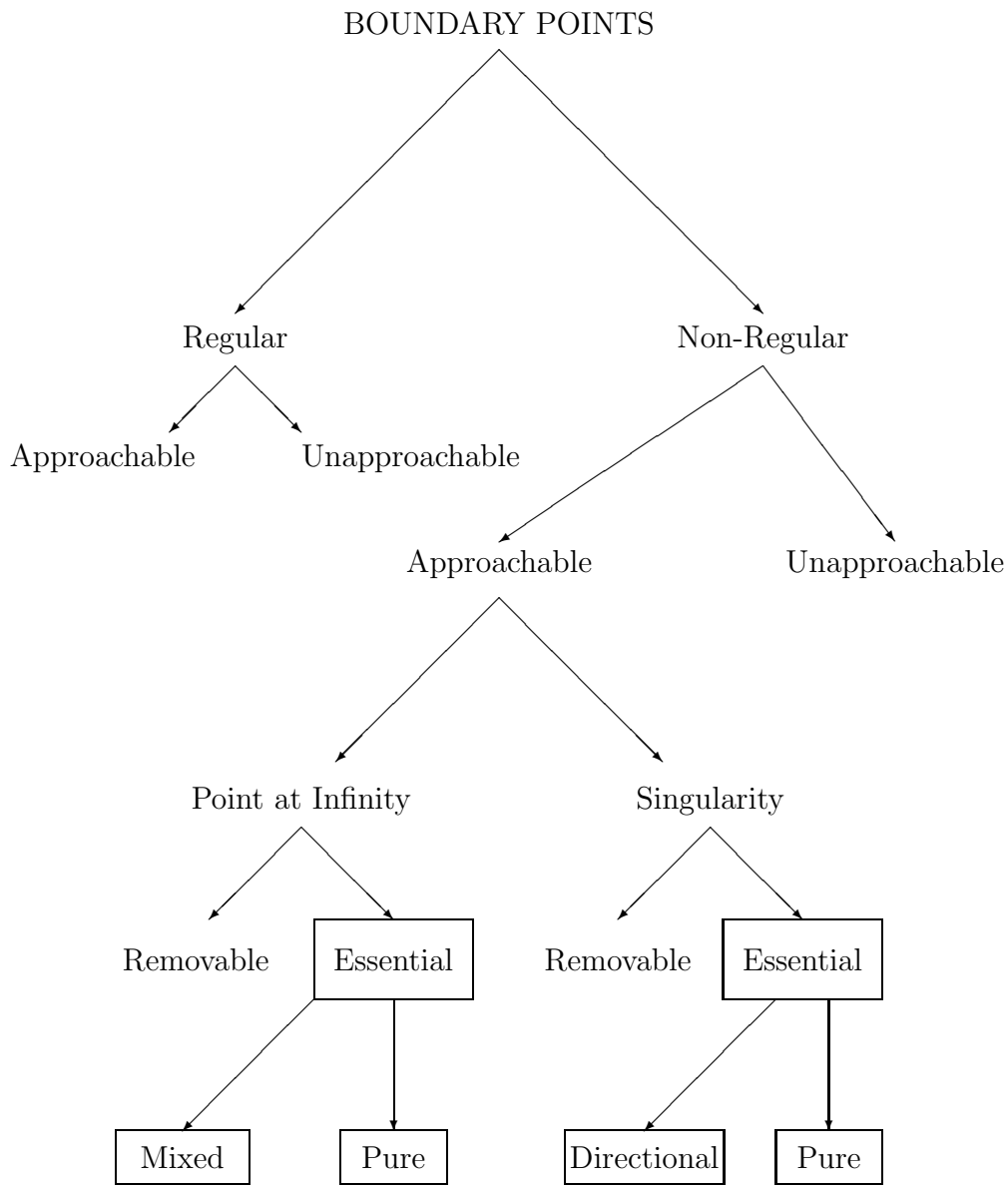
\begin{figure}
\vbox{\kern-15pt
\advance\hsize by1cm \noindent
\begin{picture}(400,500)
\put(200,480){\mb{BOUNDARY POINTS}}
\put(200,470){\vector(-1,-1){100}}
\put(200,470){\vector(1,-1){100}}
\put(100,360){\mb{Regular}}
\put(300,360){\mb{Non-Regular}}
\put(100,350){\vector(-1,-1){25}}
\put(100,350){\vector(1,-1){25}}
\put(300,350){\vector(-3,-2){100}}
\put(300,350){\vector(3,-4){50}}
\put(85,315){\mb[r]{Approachable}}
\put(125,315){\mb[l]{Unapproachable}}
\put(200,275){\mb{Approachable}}
\put(350,275){\mb{Unapproachable}}
\put(200,262){\vector(-1,-1){62}}
\put(200,262){\vector(1,-1){62}}
\put(125,185){\mb{Point at Infinity}}
\put(280,185){\mb{Singularity}}
\put(125,175){\vector(-1,-1){35}}
\put(125,175){\vector(1,-1){35}}
\put(280,175){\vector(-1,-1){35}}
\put(280,175){\vector(1,-1){35}}
\put(85,125){\mb{Removable}}
\put(245,125){\mb{Removable}}
\put(130,110){\framebox(60,30){Essential}}
\put(285,110){\framebox(60,30){Essential}}
\put(130,110){\vector(-1,-1){50}}
\put(160,110){\vector(0,-1){50}}
\put(305,110){\vector(-1,-1){50}}
\put(325,110){\vector(0,-1){50}}
\put(225,40){\framebox(60,20){Directional}}
\put(55,40){\framebox(50,20){Mixed}}
\put(140,40){\framebox(40,20){Pure}}
\put(305,40){\framebox(40,20){Pure}}
\end{picture}
\par
\kern-15pt
}
\caption{\label{fig:bp} Schematic classification of boundary points. Boxes
surround concepts which are invariant under boundary point equivalence and
therefore pass to the abstract boundary.}
\end{figure}

The whole classification as it emerges from this sequence of questions is
displayed schematically in Figure~\ref{fig:bp}.
Boxes surround concepts which pass to the abstract boundary.

\subsection*{Covering Relations} \vsp
There are eight principal categories of boundary points. These are the
regular boundary points (approachable and unapproachable regular boundary
points being regarded
as subcategories); non-regular unapproachable boundary points; removable,
mixed and pure
points at infinity; removable, directional (mixed) and pure singularities.
It is of interest to know which of these categories can or cannot cover each
other. In Table~\ref{tab:1} we put a $\surd$ if a boundary point of type
corresponding to the row label can cover a boundary point of the type belonging
to the column label (i.e., if an explicit example of such a covering can be
found), while we put a $\times$ there if it is impossible. Most of the
positions
in the table are easy to fill in, although specific examples of a covering
where
it is possible can be a little tricky to find in some cases.

The value of the table is that it allows us to see at once which pairs of
categories
can or cannot have representatives which are equivalent to each other. This
is particularly valuable when it comes to analysing the abstract boundary.
\newcommand{\iy}{$\infty$}
\newcommand{\tk}{$\surd$}
\newcommand{\x}{$\times$}

\begin{table}
\begin{tabular}{|l|c|c|c|c|c|c|c|c|} \hline
                   & reg & non-reg & rem & mix & pure & rem  & dir  & pure
\\[-1ex]
                   &     & unapp   & \iy & \iy & \iy  & sing & sing & sing
\\ \hline
regular            & \tk & \tk     & \tk & \x  & \x   & \tk  & \x   & \x \\
\hline
non-reg. unapp.    & \tk & \tk     & \x  & \x  & \x   & \x   & \x   & \x \\
\hline
remov. pt. \iy     & \tk & \tk     & \tk & \x  & \x   & \x   & \x   & \x \\
\hline
mixed pt. \iy      & \tk & \tk     & \tk & \tk & \tk  & \x   & \x   & \x \\
\hline
pure pt. \iy       & \x  & \tk     & \tk & \x  & \tk  & \x   & \x   & \x
\\ \hline
remov. sing.       & \tk & \tk     & \tk & \tk & \tk  & \tk  & \x   & \x
\\ \hline
dir. sing.         & \tk & \tk     & \tk & \tk & \tk  & \tk  & \tk  & \tk
\\ \hline
pure sing.         & \x  & \tk     & \x  & \x  & \x   & \tk  & \x   & \tk
\\ \hline
\end{tabular}
\caption{\label{tab:1} Covering table. A \tk\ means that a row-labelled
boundary
point can cover a column-labelled one, \x\ means this is impossible. The
various
labels are abbreviations for regular; non-regular unapproachable; removable,
mixed and pure points at infinity; removable, directional and pure
singularities. }
\end{table}

\subsection*{Classification of the Abstract Boundary} \vsp
We are now in a position to completely classify the abstract boundary points.
First of all abstract boundary points can be divided into approachable (\cbps)
and unapproachable. We essentially discard the latter, although there is an
interesting subclass of unapproachable abstract boundary points which have
a regular boundary point representative (Example~\ref{ex:5}). Focussing
attention on
approachable boundary points, we have already seen that the
classes which belong to the essential categories, namely mixed and pure points
at infinity and directional and pure singularities, all pass to the abstract
boundary. To see that these are the only categories which pass to the abstract
boundary one makes use of Table~\ref{tab:1}.

A category $C$ passes to the abstract boundary if and only if representative
boundary points from $C$ can {\em never} be equivalent to points from another
category $C'$.  This will be true if there is a \x\ in either the $CC'$ or
$C'C$ entry of the table for every category $C'\neq C$.  It is easily
verified that the only categories
for which this holds are those mentioned above. The remainder we may simply
term ``indeterminate''---these comprise abstract boundary points which have as
members regular boundary points, removable points at infinity and removable
singularities.  As a regular boundary point can be equivalent either to a
removable point at infinity or to a removable singularity (but never to
both) it is
not possible to create genuine subcategories of the indeterminate abstract
boundary points. In view of all this it is reasonable to make the following
definitions.

\begin{definition} {\em An abstract boundary point will be termed an {\em
abstract point at
infinity} if it has a representative boundary point which is an essential
point at infinity.}
\end{definition}

\begin{definition} {\em An abstract boundary point will be termed an {\em
abstract singularity}
if it has a representative boundary point which is an essential
singularity.} \end{definition}

\begin{figure}
\vtop{\advance\hsize by1cm
\begin{picture}(400,300)
\put(120,250){\framebox(200,30){ABSTRACT BOUNDARY POINTS}}
\put(200,250){\vector(0,-1){40}}
\put(260,250){\vector(1,-1){40}}
\put(150,180){\framebox(100,30){Approachable}}
\put(150,180){\vector(-1,-1){50}}
\put(250,180){\vector(1,-1){60}}
\put(200,180){\vector(0,-1){60}}
\put(300,180){\framebox(100,30){Unapproachable}}
\put(0,0){\framebox(100,100){}}
\put(150,60){\framebox(50,30){Mixed}}
\put(150,90){\framebox(100,30){Points at Infinity}}
\put(200,60){\framebox(50,30){Pure}}
\put(300,60){\framebox(60,30){Directional}}
\put(300,90){\framebox(100,30){Singularities}}
\put(360,60){\framebox(40,30){Pure}}
\put(0,100){\framebox(100,30){Indeterminate}}
\put(0,80){\makebox(100,20){regular}}
\put(0,60){\makebox(100,15){removable pts.}}
\put(0,45){\makebox(100,15){at infinity}}
\put(0,20){\makebox(100,15){removable}}
\put(0,5){\makebox(100,15){singularities}}
\end{picture}
}
\caption{\label{fig:abp} Classification scheme for abstract boundary points.}
\end{figure}
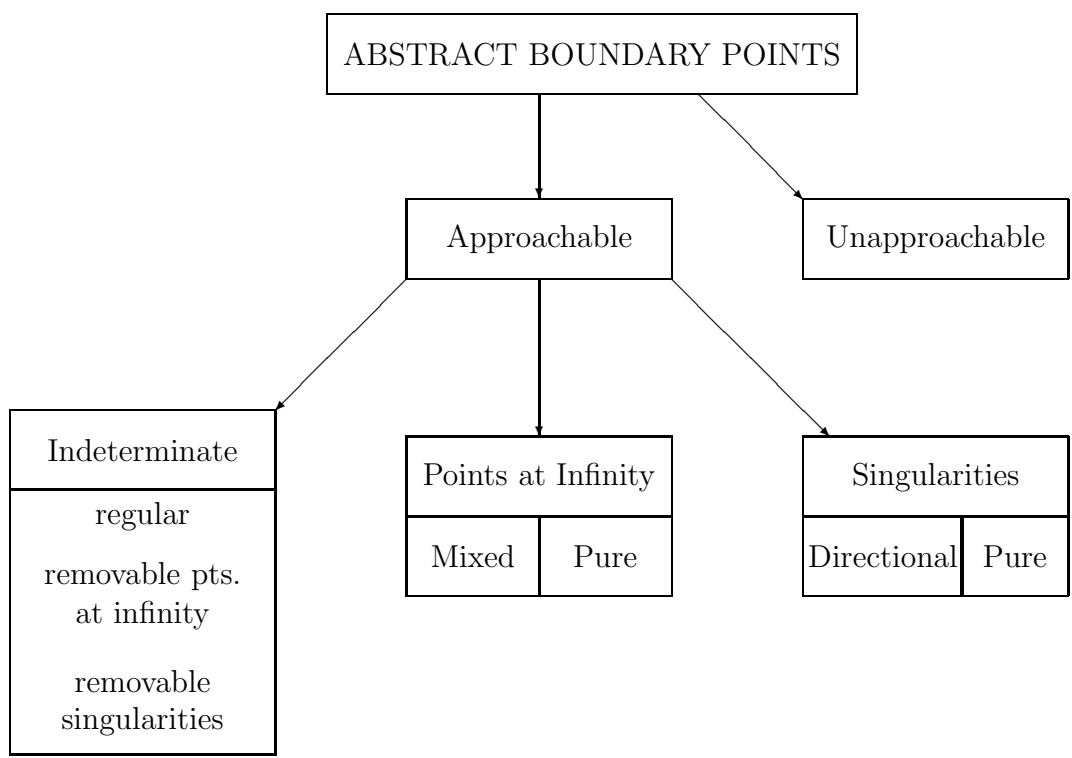

This classification together with the further subclassification into {\em
mixed} ({\em directional})
and {\em pure} classes is depicted in Figure~\ref{fig:abp}.
In this way we see that every pseudo-Riemannian manifold with a class \calc\ of
curves satisfying the b.p.p.\ has a well-defined {\em abstract
singular boundary} (consisting of the set of all abstract singularities)
and also an {\em abstract infinity}. This essentially solves the problem
originally posed by this paper, namely to construct a boundary for
an arbitrary $n$-dimensional pseudo-Riemannian manifold with class \calc\
of curves satisfying
the b.p.p.\ which represents its singularities. The abstract infinity,
representing the ``boundary
at infinity'', comes as a bonus.

\section{Singularities of pseudo-Riemannian manifolds}

\begin{definition} {\em We will say that a pseudo-Riemannian manifold
$(\calm,g,\calc)$ with class \calc\ of
curves satisfying the b.p.p.\ has a $C^{l}$ {\em
singularity\/} if its abstract $C^l$ singular boundary is non-empty, i.e.,
if there exists an envelopment of \calm\ having an
essentially $C^l$ singular boundary point.
Conversely, $(\calm,g,\calc)$ will be called $C^{l}$ {\em singularity-free\/}
if it has no $C^{l}$ singularities, i.e., if for every envelopment of
\calm\ its boundary
points are either $C^l$ non-singular ($C^l$ regular boundary points, $C^l$
points at infinity or
unapproachable boundary points) or $C^l$ removable singularities.}
\end{definition}

At first sight it might seem a daunting prospect to decide whether a given
$(\calm,g,\calc)$ is singularity-free, as it seems that one would have to
investigate
every possible envelopment of \calm\ and check whether it has any essential
singularities
or not. In practice, however, several very general theorems exist to make
the task
much simpler. The two which follow are generally regarded as the {\em sine qua
 non} of any successful theory of singularities \cite{ShepRy}.

\begin{theorem} \label{th:comp}
Every compact pseudo-Riemannian manifold $(\calm,g)$ is singularity-free
for any family \calc\ of
curves with the b.p.p.
\end{theorem}
{\em Proof:} A compact manifold has no non-trivial envelopments, for any
enveloping manifold \hM\ would contain \calm\ as a compact open subset. But
\calm\ cannot
be both open and closed, since the enveloping manifold \hM\ is assumed to
be connected. Since \calm\ has no envelopments, its abstract boundary is empty
and, in particular, can contain no singularities. \qed \vsp

\begin{theorem} \label{th:ccomp}
Every pseudo-Riemannian manifold with a family \calc\ of curves satisfying
the b.p.p.,
$(\calm,g,\calc)$, which is \calc-complete is singularity-free.
\end{theorem}
{\em Proof:} Let $(\calm,g,\calc,\hM,\phi)$ be any envelopment of
$(\calm,g,\calc)$
and let $p$ be any \cbp\ of this envelopment.
Let $\gamma:[a,b) \ra \calm$ be some curve from \calc\ having $p$ as a
limit point.
The parameter range $[a,b)$ cannot be bounded, else by \calc-completeness
$\gamma$ would have an endpoint $q\in \calm$.
Since endpoints are unique limit points and
$q\neq p$ (since $p\not\in \calm$), it is clear that this yields a
contradiction. Hence
$p$ cannot be a singularity.
\qed \vsp

A rather stronger version of this theorem is available when, as is usually
required, $\calc \supseteq \calc_g(\calm)$. It implies that not even
unapproachable
regular boundary points are possible in this case, which would still have
been permitted
by Theorem~\ref{th:ccomp}.

\begin{theorem} \label{th:rcomp}
If the pseudo-Riemannian manifold $(\calm,g)$ is $\calc_g(\calm)$-complete
then it can have no
regular boundary points. \end{theorem}
{\em Proof:} Let $(\calm,g,\hM,\phi)$ be any envelopment of $(\calm,g)$ and
let $p$ be any
boundary point of this envelopment. If $p$ is a regular boundary point then
it is possible to
find a neighbourhood \calu\ of $p$ in \hM\ on which a metric $\bar{g}$ exists
which extends $g|_{{\cal U}\cap{\cal M}}$. There is no loss of generality
in assuming that \calu\ is a normal neighbourhood. Let $q \in
\calu\cap\calm$ and
$\gamma:[a, b] \ra \calu$ be the unique geodesic for the metric $\bar{g}$
connecting $q$ to $p$. This geodesic clearly intersects \calm\ (since $q
\in \calm$)
and has bounded parameter. On the other hand it must exit \calm\ since $p
\not\in \calm$. Let the
first parameter value $t \in (a,b]$ for which $\gamma(t) \not\in \calm$ be
denoted by $c$. Clearly
the geodesic $\gamma |_{[a,c)}$ does not have its endpoint $\gamma(c)$ in
\calm. This contradicts
$\calc_g(\calm)$-completeness, whence $p$ cannot be regular. \qed \vsp

Of course it is possible for a pseudo-Riemannian manifold $(\calm,g)$ to be
$\calc_{g}(\calm)$-complete and therefore to be singularity-free, but if
the family of curves  is
extended to a wider class such as $\calc =\calc_{gap}(\calm)$, it might no
longer be \calc-complete.
A classic example of this kind has been given by Geroch \cite{Ger2}, where
a space--time
is geodesically complete but has incomplete curves of bounded acceleration.

\begin{definition} {\em Boundary points arising as limit points of curves
from the family
\calc\ with bounded parameter in a $\calc_g(\calm)$-complete manifold will
be referred to as
{\em Geroch points}.} \end{definition}

Geroch points must be singular (as is clear from Theorem~\ref{th:rcomp}). In
fact they must be essentially singular, else they would be covered by a
boundary
set of another embedding which contains at least one regular boundary point (by
Theorem~\ref{th:removs}), which is impossible by Theorem~\ref{th:rcomp}.
Thus the existence of
Geroch points implies that the pseudo-Riemannian manifold $(\calm,g,\calc)$
has a singularity.
Note, however, that if $\calc_g(\calm)$-completeness had been taken as
one's goal, then the
manifold would be singularity-free.  It is therefore of vital importance to
specify the family \calc\ of curves when discussing the question
of singularities.

\begin{figure}
	\leavevmode\centering\epsfbox{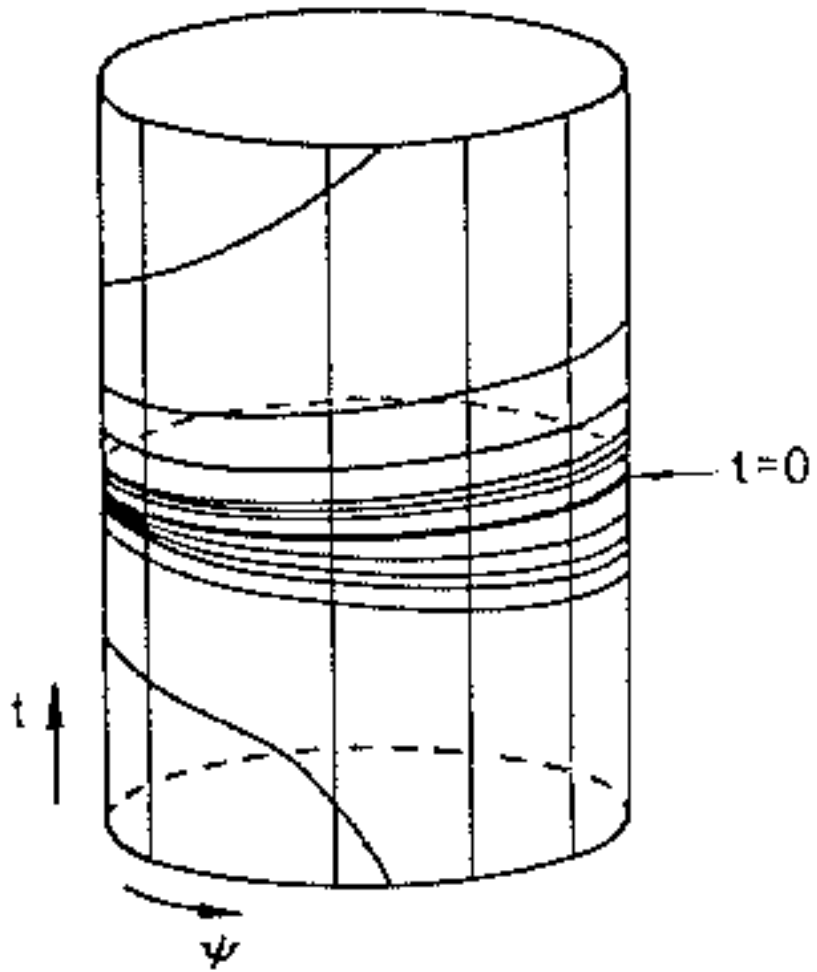}
        \caption{\label{fig:mis} Misner's example.}
\end{figure}

Also with this scheme, geodesic incompleteness does not necessarily imply
that the particular
$(\calm,g,\calc)$ in question has a singularity. The classic example of
this is the Taub--NUT
space--time \cite{HE} or Misner's simplified version \cite{Mis}, which we
present here.
\begin{example} \label{ex:mis}
{\em Let \calm\ be the 2-dimensional manifold $S^1 \times {\Bbb R}^{1}$, with
Lorentzian metric
\[{\rm d}s^2 = \,2\, {\rm d}t\, {\rm d}\psi + t\, {\rm d}\psi^2, \mbox{
where }\: t\in {\Bbb R},\: 0\leq\psi< 2\pi\,.\] The central circle $t=0$ and
the vertical
lines $\psi = $ constant are
complete null geodesics, but there are other geodesics (null, timelike
and spacelike) which execute infinite spirals as they approach $t=0$
from either above or below (see Figure~\ref{fig:mis}).  These geodesics all
approach $t=0$ with bounded affine parameter and thus are either past-
or future-incomplete. On the other hand it is clear that there is no
envelopment of this space--time providing  boundary points which are
limit points of these incomplete curves (this is seen most readily
by compactifying the space into a torus by identifying $t=-\infty$
with $t=+\infty$, for then no envelopments exist at all, as was shown in
Theorem~\ref{th:comp}). Hence this space--time is singularity-free but is
geodesically incomplete.
In many sources \cite{ShepRy,ES} this space--time is classified as singular.
This interpretation seems to arise in part because more than one extension
is possible across $t=0$.  Thus an extension of the lower half-space
($t<0$) exists in which the spiralling geodesics are complete, but the
vertical ones become incomplete spirals.  Undesirable as this sort of
behaviour may be physically, we do not see it as grounds for calling the
space--time singular. Indeed, there can be other reasons apart from
singularities
for discarding a metric on physical grounds---for example, the existence
of closed causal curves. These, incidentally, are also present in the
Misner space--time.}
\end{example}

\section{Conclusions}
We have presented a new definition of singularities which can be applied
equally well
to manifolds of any dimension and metric of any signature. The key idea has
been
to define the {\em abstract boundary\/} or {\em a-boundary} \bm\ of a manifold
\calm. This is definable for any manifold whatsoever and includes, in a
sense, all
possible boundary points which can arise from open embeddings of the manifold.
The {\em a}-boundary is constructed entirely from the manifold itself
and is therefore something which every
manifold gets {\em gratis}. When the manifold is endowed with extra structure,
such as a pseudo-Riemannian metric or an affine connection, then the
approachable boundary
points can be classified into three important categories---regular boundary
points,
points at infinity and singularities, together with further subcategories. The
key to this classification is the specification of a family of parametrised
curves in
the manifold satisfying the bounded parameter property. This is a vital
ingredient, for different such families will give rise to different singularity
structures. It is usual, however, to insist that the family does include all
affinely parametrised geodesics.

The scheme presented is, we believe, very robust and passes most standard
tests required of a theory of singularities. Furthermore, it is a practical
scheme, for when a pseudo-Riemannian manifold such as a space--time is
presented, it is usually
given in a specific coordinate system. This often amounts to giving an
envelopment of the manifold in question.
It is normally then a relatively straightforward matter to classify
the boundary points (and, by equivalence, the abstract boundary points of which
they are representatives) arising from this envelopment.
Frequently the information so obtained is sufficient to obtain an analysis of
the {\em a}-boundary which at least suffices for answering the main
questions about
the singularity structure of the particular pseudo-Riemannian manifold.
Some examples can be
found in ref.~\cite{Scott} and others will be given in a forthcoming paper.

One of the great benefits of our scheme is that when a closed region is excised
from a singularity-free pseudo-Riemannian manifold, the resulting
pseudo-Riemannian manifold is
still singularity-free, since only regular boundary points are introduced
by the
excision process.  It was never possible to make such a claim with previous
schemes because geodesics always had to be maximally extended before the
discussion could begin. As maximal extensions of pseudo-Riemannian
manifolds are not easy to
find and are not even unique in the analytic case, we believe this to be a
great
advantage of our approach.

Finally, a number of questions remain unanswered in this paper. In
particular, no mention has been made of the topology of the {\em a}-boundary,
especially its singular part. This is an important topic, which we propose
to discuss in another paper. It would be of great interest to know
how the {\em a}-boundary and its topology relates to the Cauchy completion
in the
case of a Riemannian manifold. Another interesting question is the
following: does every essential singularity cover a pure singularity? In
other words, is there
always a ``pure core'' to the singular part of the {\em a}-boundary? For
any envelopment of a
pseudo-Riemannian manifold, is every boundary point coverable by a boundary
set of some
embedding, all of whose boundary points are approachable by geodesics? Does
every manifold
\calm\ have an envelopment such that its closure \bM\ is compact, i.e.,
does \calm\ have a
compactification? Obviously one could go on, but despite the interest of
such questions there
is nothing in them to negate the consistency and completeness of our scheme.

A question of particular interest would be to see how the {\em a}-boundary
relates
to other boundary constructions such as the {\em b}-boundary or {\em
g}-boundary.  As it
stands it is difficult to see any connection, at least until the further topic
of topology on the boundary is addressed. Our main objective in this paper
has been to answer the question {\em when is a manifold with affine
connection and preferred family of curves singular?} The {\em a}-boundary
seems to be a
satisfactory vehicle for dealing with this question. The criteria we have
arrived at are
unambiguous and in many cases can be readily shown to give the expected answer
(see ref.~\cite{Scott} for applications to such examples as Schwarzschild,
Friedmann, Curzon, etc.).
To explicitly display the {\em a}-boundary of a given manifold, however,
is not a feasible proposition in general, since it amounts to specifying
every inequivalent way in which the manifold can be embedded as an open
submanifold of a larger manifold. Because of the prevalence of ``blow-up''
maps, this is clearly not a practical thing to do. Some method for cutting
down or ``sectioning'' the {\em a}-boundary into manageable sized portions will
be needed before structures such as topology can be attached. We
will present procedures for doing this in a forthcoming publication.

\newpage
\subsubsection*{Acknowledgements}
P.S. would like to express his appreciation to the Department
of Applied Mathematics, University of Waterloo and particularly to R.G.
McLenaghan for their hospitality during some of this work.
S.M.S. would like to express her appreciation to the Aspen Center for Physics,
Colorado, where some of this work was also carried out.
We have benefitted greatly from discussions with C.J.S. Clarke, R. Geroch,
R. Penrose, W. Unruh and J. Vickers. We would also like to thank the
referee B.G. Schmidt for
some helpful suggestions regarding the readability of the paper, as well as
members
of the Mathematical Relativity Group at the Australian National University for
carefully checking the manuscript. 

\end{document}